\documentclass[11pt,a4paper,conference]{IEEEtran}

\usepackage{graphicx}
\usepackage{amssymb}
\usepackage{caption}
\usepackage{subcaption}
\usepackage{url}
\usepackage{amsmath}
\usepackage{float}
\usepackage{caption}
\usepackage{color}

\graphicspath{{./Images/}}

\begin{document}

%
%

\title{Bifurcations in valveless pumping techniques from a coupled fluid-structure-electrophysiology model in heart development}

%
%

\author{\IEEEauthorblockN{Nicholas A. Battista}
\IEEEauthorblockA{Dept. of Mathematics and Statistics\\
The College of New Jersey\\
Ewing, NJ, USA\\
battistn@tcnj.edu}
\and
\IEEEauthorblockN{Laura A. Miller}
\IEEEauthorblockA{Dept. of Biology, Dept. of Mathematics\\
University of North Carolina at Chapel Hill\\
Chapel Hill, NC, USA\\
lam9@unc.edu}
} 

\maketitle

%
%

\begin{abstract}
We explore an embryonic heart model that couples electrophysiology and muscle-force generation to flow induced using a $2D$ fluid-structure interaction framework based on the immersed boundary method. The propagation of action potentials are coupled to muscular contraction and hence the overall pumping dynamics. In comparison to previous models, the electro-dynamical model does not use prescribed motion to initiate the pumping motion, but rather the pumping dynamics are fully coupled to an underlying electrophysiology model, governed by the FitzHugh-Nagumo equations. Perturbing the diffusion parameter in the FitzHugh-Nagumo model leads to a bifurcation in dynamics of action potential propagation. This bifurcation is able to capture a spectrum of different pumping regimes, with dynamic suction pumping and peristaltic-like pumping at the extremes. We find that more bulk flow is produced within the realm of peristaltic-like pumping.
\end{abstract}

\begin{IEEEkeywords}
valveless pumping; heart development; immersed boundary method; fluid-structure interaction; mathematical biology; biomechanics
\end{IEEEkeywords}

%
%

\section{Introduction}
\label{intro}

Various kinds of hearts are found throughout the animal kingdom \cite{Alters:2000,Xavier:2007,Vivien:2016}. In particular many invertebrates have valveless, tubular hearts from their infancy throughout adulthood \cite{Calabrese:2014}. These tubular hearts are similar to vertebrate heart morphologies during their first stage of vertebrate heart morphogenesis, e.g., the linear heart tube stage. We begin our discussion of heart tube morphologies by considering the evolution of hearts in the animal kingdom.

\begin{figure}
    \centering
    \includegraphics[width=0.5\textwidth]{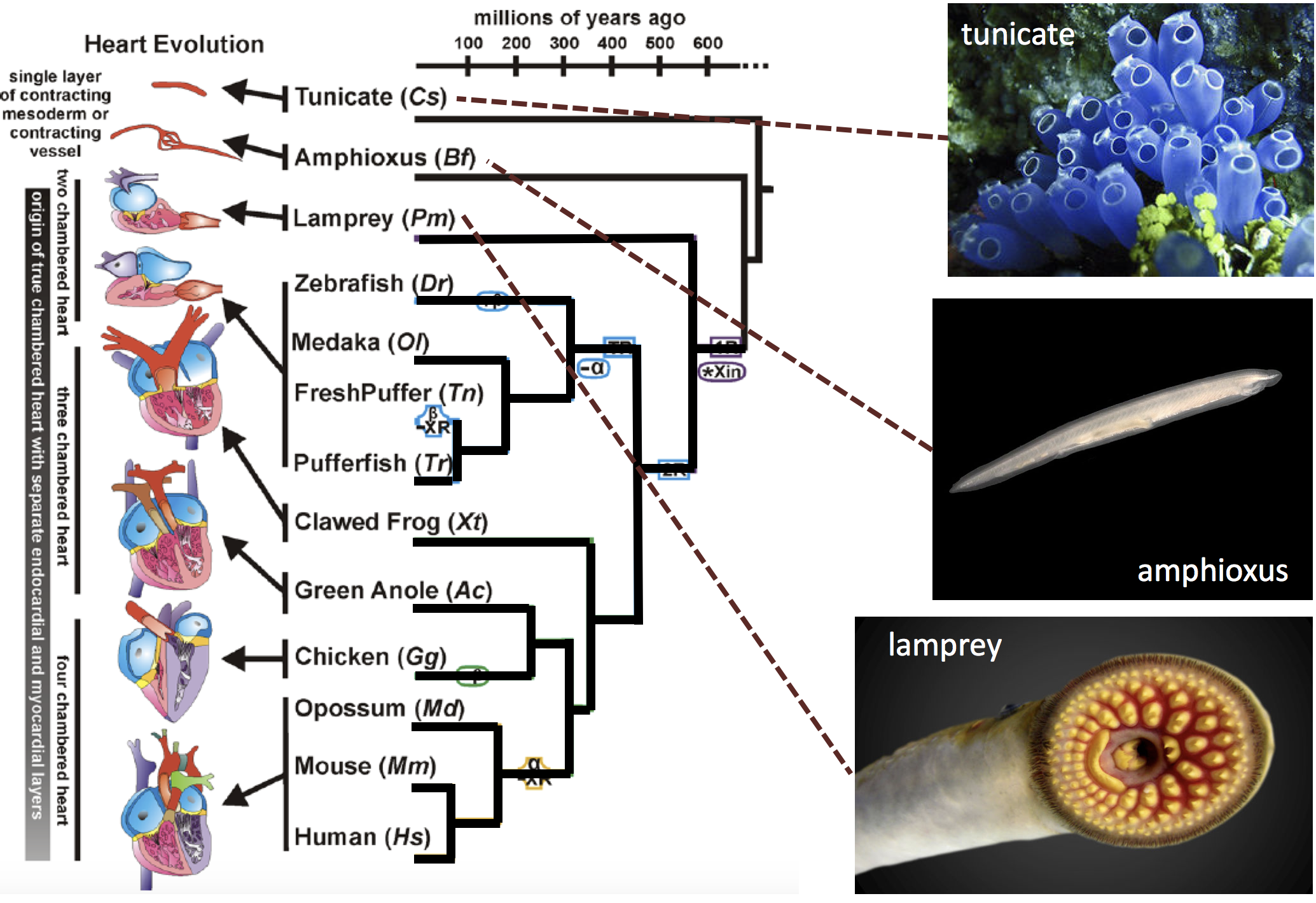}
    \caption{Figure adapted from Grosskurth \textit{et al.} \cite{Grosskurth:2008} illustrating the evolution of hearts from the valveless heart tubes in the open circulatory systems of tunicates to the adult multi-chambered-valvular of vertebrates. Tunicate, amphioxus, and lamprey images adapted from \cite{TunicateImage,LanceletImage,LampreyImage}, respectively.}
    \label{fig:HT_Heart_Evo}
\end{figure}

Figure \ref{fig:HT_Heart_Evo} shows the evolution of hearts from tunicates to humans. Tunicates have an open circulatory system from infancy through adulthood, in which blood is pushed through out the organism by a valveless-tubular heart \cite{Alters:2000,Konrad:2016}, which is composed of only a single layer of myocardial cells. Next on the evolutionary chain is the amphioxus. The amphioxus heart is a rostrocaudally extended tube from its infancy through adulthood \cite{Holland:2003}. Similar to the tunicate heart, an amphioxus heart consists only of a monolayer of myocardial cells. Furthermore its heart has no chambers, valves, endocardium, epicardium, or other differentiated features of vertebrate hearts. Still, the amphioxus is regarded as the closest living invertebrate relative to vertebrates \cite{Holland:2001} and appears fish-like. 

\begin{figure}
    \centering
    \includegraphics[width=0.4\textwidth]{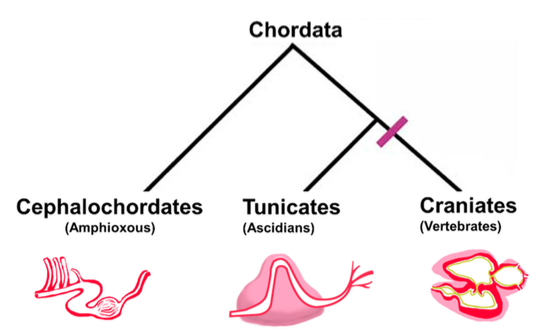}
    \caption{Figure illustrating the phylogenetic relationship and general heart structure of the Chordate subphyla. Cephalochordates, like amphioxus, have a series of four peristaltic vessels that serve as a pump, while tunicates have a single-chamber pump, which is composed of a single layer of myocardium (red) surrounded by stiff pericardial layer (pink). The earliest vertebrates, e.g., lampreys, have at least a two-chambered myocardium composed of a layer of cardiac myocardial cells (red), an endocardial cellular layer (yellow), valves that separate distinct chambers, and a surrounding pericardium (pink). Figure adapted from \cite{Anderson:2016}.}
    \label{fig:Heart_Evo_Cartoon}
\end{figure}

Furthermore, Figure \ref{fig:HT_Heart_Evo} illustrates the bifurcation to multi-chambered hearts in a vertebrate - the lamprey. Lampreys are jawless fishes that are a very ancient lineage of vertebrates \cite{Smith:2015}. The lamprey is considered to have four heart chambers, which are the sinus venosus, atrium, ventricle, and conus arteriosus \cite{Percy:1986}. This is similar as to the zebrafish heart, which contains four chambers - the sinus venosus, atrium, ventricle, and bulbus arteriosus. Lamprey hearts also are valvular pumping systems, containing valve leaflets between chambers \cite{Green:2015}. Moreover, lampreys are the first organism to develop an endocardial layer in addition to a myocardium, as well as, the first organism to develop cardiac valves \cite{Puceat:2012}. An evolutionary depiction of heart morphology is illustrated in Figure \ref{fig:Heart_Evo_Cartoon}, which was adapted from \cite{Anderson:2016}.

However, as discussed, the vertebrate embryonic heart is a valveless tube, similar to those in various invertebrates, such as urochordates and cephalochordates \cite{Kriebel:1967,Randall:1980}, making invertebrates like sea squirts a possible model for heart development \cite{Morad:1980}. Historically, the pumping mechanism in these hearts has been described as peristalsis \cite{Kriebel:1967, Santhanakrishnan:2011}, while more recently, dynamic suction pumping (DSP) has been proposed as a novel cardiac pumping mechanism for the vertebrate embryonic heart by Kenner \emph{et. al.} in $2000$ \cite{Kenner:2000}, and later declared the main pumping mechanism in vertebrate embryonic hearts by Forouhar \emph{et. al.} in $2006$ \cite{Forouhar:2006}. Debate over which is the actual pumping mechanism of the embryonic heart continues today, with the possibility that the mechanism may vary between species or may be some hybrid of both mechanisms \cite{Waldrop:15BMMB,Manner:2010}.  

The Liebau pump, a dynamic suction pump, was first described in 1954 \cite{Meier:2011}, and was studied as a novel way to pump water. It has not been until the past 20 years that scientists started looking at the pump as a valveless pumping mechanism in many biological systems and biomedical applications, including microelectromechanical systems (MEMs) and micro-fluidic devices. Direct applications of such pumps include tissue engineering, implantable micro electrodes, and drug delivery \cite{Lee:2004,Chang:2007,Lee:2008,Meier:2011}. 

\begin{figure}[H]
    \centering
    \includegraphics[width=0.45\textwidth]{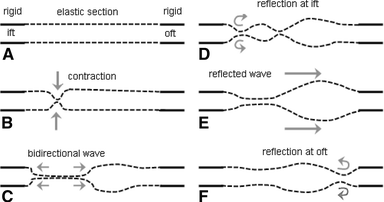}
    \caption{Schematic diagram illustrating dynamic suction pumping \cite{Santhanakrishnan:2011}. (A) The flexible tube is at rest. (B) Active contraction of the tube in a non-central location along the tube. (C) Contraction induces an elastic passive bidirectional wave to propagate along the tube. (D) Wave reflects off rigid portion of the tube on side nearest to contraction point. (E) The reflected wave travels down the tube. (F) The waves reflect off the rigid section at the far side of the tube. Notice the the reflected wave amplitude is smaller than the reflected wave off the other end.}
    \label{fig:DSP_Schematic}
\end{figure}

With extensive industrial applications, dynamic suction pumping has proven to be a suitable means of transport for fluids and other materials in a valveless system, for scales of $Wo>1$ \cite{Baird:2014}. DSP can be most simply described by an isolated region of actuation, located asymmetrically along a flexible tube with stiffer ends. Flexibility of the tube is required to allow passive elastic traveling waves, which augment bulk transport throughout the system. The rigid ends of the tube cause the elastic waves to reflect and continue to propagate in the opposite direction, which when coupled with an asymmetric actuation point, can promote unidirectional flow. DSP is illustrated in Figure \ref{fig:DSP_Schematic}.

Due to a coupling between the system's geometry, material properties of the tube wall, and pumping mechanics, there is a complex, nonlinear relationship between volumetric flow rate and pumping frequency \cite{Baird:2014,Bringley:2008,Hickerson:2005}. Analytic models of DSP have been developed to address this relationship \cite{Ottsen:2003,Auerbach:2004,Manopoulos:2006,Samson:2007,Bringley:2008,Babbs:2010}. Most models use simplifications, such as an inviscid assumption, long wave approximation, small contraction amplitude, and/or one-dimensional flow. Furthermore, no analytical model has described flow reversals, which can occur with changes in the pumping frequency. Relaxing many of these assumptions, physical experiments have been performed to better understand DSP \cite{Hickerson:2005,HickersonThesis:2005,Bringley:2008,Meier:2011}, as well as \emph{in silico} investigations \cite{Jung:1999,Jung:2001,Avrahami:2008,Baird:2014,BairdThesis:2014,Battista:2016a}. Most of the joint experimental and computational studies focus on the `high' $Wo$ regime ($Wo>>1$), besides studies by Baird \textit{et al.} \cite{BairdThesis:2014}.

\begin{figure}[H]
  \centering
  \includegraphics[width=0.4\textwidth]{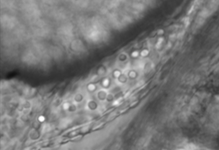}
  \caption{The embryonic heart tube of a Zebrafish 30 hpf, courtesy of \cite{Maes:2011}. Spherical blood cells are seen within the tubular heart. The heart tube is roughly $5$ blood cells thick in diameter.}
  \label{Maes30hpf}
\end{figure}

In this paper we will investigate the pumping phenomena that occurs as a result of a coupled fluid-structure-electrophysiology model \cite{Baird:2015}, and the bulk flow rates thereby induced. The electrophysiology model governs the propagation of action potentials, which then are coupled to muscular contraction, and hence the overall pumping dynamics. We then perturb the diffusion parameter in the electrophysiology model to investigate the bifurcation in pumping dynamics that occurs as a result of differing action potential propagation. This bifurcation is able to capture a spectrum of different pumping regimes, i.e., dynamic suction pumping to peristaltic-like waves of contraction. The electrophysiology model is governed by the FitzHugh-Nagumo equations \cite{FitzHugh:1961,Baird:2015}.

%
%

\section{Methods}
\label{sec:methods}

The immersed boundary method (IBM) is a numerical method developed to solve problems involving viscous, incompressible fluid coupled to the movement of an immersed elastic structure \cite{Peskin:2002,Mittal:2005,BattistaIB2d:2016}. Since its development in the 1970s by Charles Peskin \cite{Peskin:1977}, it has been applied to a wide spectrum of biomathematical models, ranging from blood flow through the heart \cite{Peskin:1977,Peskin:2002}, aquatic locomotion \cite{Hieber:2008,Hoover:2015}, insect flight \cite{Miller:2004,Miller:2009}, to plant biomechanics \cite{Miller:2012,BattistaIB2d:2017}, and muscle mechanics \cite{Battista:2015,BattistaIB2d:2016}. 

The power of this method is that it can be used to describe flow around complicated time-dependent geometries using a regular Cartesian discretization of the fluid domain. The elastic fibers describing the structure are discretized on a moving curvilinear mesh defined in the Lagrangian frame. The fluid and elastic fibers constitute a coupled system, in which the structure moves at the local fluid velocity and the structure applies a singular force of delta-layered thickness to the fluid. 

%
%

\subsection{Equations of the IBM}
Assume that the immersed boundary is described on a curvilinear, Lagrangian mesh, $S$, that is free to move. The fluid is described on a fixed Cartesian, Eulerian grid, $\Omega$, that has periodic boundary conditions. Given the size of the domain and the localization of the flow to the tube, the boundary conditions do not significantly affect the fluid motion. The governing equations for the fluid, the Navier-Stokes equations, are given by
\begin{align}
\nonumber \rho \left[ \frac{\partial {\bf{u}} }{\partial t}( {\bf{x}},t ) +  {\bf{u}}( {\bf{x}},t )\cdot \nabla {\bf{u}}( {\bf{x}},t ) \right] &= -\nabla p( {\bf{x}},t )\\
\label{momentum} + \mu \Delta&{\bf{u}}( {\bf{x}},t ) + {\bf{f}}( {\bf{x}},t ) 
\end{align}
\begin{equation}
    \label{incompressible} \nabla\cdot{\bf{u}}({\bf{x}},t) = 0.
\end{equation}
Eqs.(\ref{momentum}) and (\ref{incompressible}) are the Navier-Stokes equations written in Eulerian form, where Eq.(\ref{momentum}) is the conservation of momentum for a fluid and Eq.(\ref{incompressible}) is the conservation of mass, i.e., incompressibility condition. The two constant parameters in these equations are the fluid density, $\rho$, and the dynamic viscosity of the fluid, $\mu$. The fluid velocity, ${\bf{u}}( {\bf{x}},t )$, pressure, $p( {\bf{x}},t )$, and body force, ${\bf{f}}( {\bf{x}},t)$, are unknown spatial time-dependent functions of the Eulerian coordinate, ${\bf{x}}$, and time, $t$. The body force describes the transfer of momentum onto the fluid due to the restoring forces arising from deformations of the elastic structure. It is this term, ${\bf{f}}( {\bf{x}},t)$, that is unique to the particular model being studied. 

The material properties of the structure may be modeled to resist to bending, stretching, and displacement from a tethered position. Other forces that can have been modeled include the action of virtual muscles, electrostatic (contact) forces, molecular bonds, porosity, and other external forces \cite{Peskin:2002,Tytell:2010,Mahur:2012,Fogelson:2008,Battista:2015,BattistaIB2d:2016}. The immersed structure may deform due to bending forces and/or stretching and compression forces. In this paper, elastic forces are calculated as beams that may undergo large deformations and Hookean springs, i.e.,
{\footnotesize{
\begin{align}
\label{beam_force} \mathbf{F}_{beam}&=-k_{beam} \frac{\partial^4}{\partial s^4}\Big( \mathbf{X}(s,t) - \mathbf{X}_B(s) \Big) \\
\label{spring_force} \mathbf{F}_{spring}&=-k_{spring} \left( 1 - \frac{R_L}{\left|\left| \mathbf{X}_{S} - \mathbf{X}_M \right|\right| } \right) \cdot \left( \mathbf{X}_M - \mathbf{X}_S \right).
\end{align}
}}
Eq.(\ref{beam_force}) is the beam equation, which describes forces arising from bending of the elastic structure. Eq.(\ref{spring_force}) describes the force generated from stretching and compression of the structure. The parameters, $k_{beam}$ and $k_{spring}$, are the stiffness coefficients of the beam and spring, respectively, and $R_L$ is the resting length of the Hookean spring. The variables $\mathbf{X}_M$ and $\mathbf{X}_S$ give the positions in Cartesian coordinates of the master and slave nodes in the spring formulations, respectively, $\mathbf{X}_B(s)$ describes the deviation from the preferred curvature of the structure. In all simulations, $\mathbf{X}_B(s) = 0$ along the straight portion of the tube.

A target point formulation can be used to tether the structure or subset thereof in place, holding the Lagrangian mesh in a preferred position that may be time dependent. An immersed boundary point with position $\mathbf{X}(s,t)$ that is tethered to a target point, with position $\mathbf{Y}(s,t)$ undergoes a penalty force that is proportional to the displacement between them. The force that results is given by the equation for a linear spring with zero resting length, 
\begin{equation}
\label{target_force} \mathbf{F}_{target}= -\kappa_{target}\left( {\bf{X}}(s,t) - {\bf{Y}}(s,t) \right), \\
\end{equation}
where $k_{target}$ is the stiffness coefficient of the target point springs. $k_{target}$ can be varied to control the deviation allowed between the actual location of the boundary and its preferred position. The total deformation force that will be applied to the fluid is a sum of the above forces, 
\begin{equation}
\label{deform_force} {\bf{F}}(s,t)= \mathbf{F}_{spring} + \mathbf{F}_{beam} + \mathbf{F}_{target}\\
\end{equation}

A more detailed description of existing fiber models can be found in \cite{BattistaIB2d:2016}. Once the total force from Eq.(\ref{deform_force}) has been calculated, it needs to be spread from the Lagrangian frame to the Eulerian grid. This is achieved through an integral transform with a delta function kernel,  
\begin{equation}
\label{body_force} {\bf{f}}({\bf{x}},t)= \int \bf{F}(s,t) \delta({\bf{x}}-{\bf{X}}(s,t)) ds. \\ 
\end{equation}
Similarly, to interpolate the local fluid velocity onto the Lagrangian mesh, the same delta function transform is used, 
\begin{equation}
\label{struc_velocity} {\bf{U}}(s,t)  = \frac{\partial {\bf{X}} }{\partial t}(s,t) =  \int {\bf{u}}( {\bf{x}}, t) \delta ({\bf{x}}-{\bf{X}}(s,t)) d{\bf{x}}. \\
\end{equation}
Eqs.(\ref{body_force}) and (\ref{struc_velocity}) describe the coupling between the immersed boundary and the fluid, e.g., the communication between the Lagrangian framework and Eulerian framework. The delta functions in these equations make up the heart of the IBM, as they are used to spread and interpolate dynamic quantities between the fluid grid and elastic structure, e.g., forces and velocity. The quantity ${\bf{X}}(s,t)$ gives the position in Cartesian coordinates of the elastic structure at local material point, $s$, and time $t$. In approximating these integral transforms, a discretized and regularized delta function, $\delta_h(\mathbf{x})$ \cite{Peskin:2002}, is used, 
\begin{equation}
\label{delta_h} \delta_h(\mathbf{x}) = \frac{1}{h^2} \phi\left(\frac{x}{h}\right) \phi\left(\frac{y}{h}\right),
\end{equation}
where $\phi(r)$ is defined as

\begin{equation}
\label{delta_phi} \phi(r) = \left\{ \begin{array}{c} \frac{1}{4}\left[1 + \cos\left(\frac{\pi r}{2}\right)\right] \ \ \ \ \ |r|\leq 2 \\
\mbox{ } \\
\ \ \ \ \ \ \ \ \ \ 0 \ \ \ \ \ \  \ \ \ \ \ \ \ \  \mbox{otherwise}. \end{array} \right.
\end{equation}

%
%
%
\subsection{Numerical Algorithm}
As stated above, we impose periodic boundary conditions on the rectangular domain. To solve Eqs. (\ref{momentum}), (\ref{incompressible}),(\ref{body_force}) and (\ref{struc_velocity}) we need to update the velocity, pressure, position of the boundary, and force acting on the boundary at time $n+1$ using data from time $n$. IBM does this in the following steps \cite{Peskin:2002}:
\indent\textbf{Step 1:} Find the force density, ${\bf{F}}^{n}$ on the immersed boundary, from the current boundary configuration, ${\bf{X}}^{n}$.\\
\indent\textbf{Step 2:} Use Eq.(\ref{body_force}) to spread this boundary force from the curvilinear mesh to nearby fluid lattice points.\\
\indent\textbf{Step 3:} Solve the Navier-Stokes equations, Eqs.(\ref{momentum}) and (\ref{incompressible}), on the Eulerian domain. In doing so, we are updating ${\bf{u}}^{n+1}$ and $p^{n+1}$ from ${\bf{u}}^{n}$ and ${\bf{f}}^{n}$. Note: because of the periodic boundary conditions on our computational domain, we can easily use the Fast Fourier Transform (FFT) \cite{Cooley:1965,Press:1992}, to solve for these updates at an accelerated rate.\\
\indent\textbf{Step 4:}  Update the material positions, ${\bf{X}}^{n+1}$,  using the local fluid velocities, ${\bf{U}}^{n+1}$ with ${\bf{u}}^{n+1}$ and Eq.(\ref{struc_velocity}).

The above steps outline the process used by the IBM to update the positions and velocities of both the fluid and elastic structure. A more detailed discussion of the IBM can be found in \cite{Peskin:2002}.

%
%

\subsection{Computational Model}

We numerically model a $2D$ closed racetrack where the walls of the tube are modeled as $1D$ fibers. The closed tube is composed of two straight portions, of equal length, connected by two half circles, of equal inner and equal outer radii. The tube, or racetrack, has uniform diameter throughout. This is similar to the racetrack model geometry as in \cite{Baird:2015}. Furthermore, as in \cite{Baird:2015}, we include the presence of an idealized stiff pericardium surrounding the flexible region of the heart tube. 

The tunicate heart consists of a myocardium which is surrounded by a stiff pericardium \cite{Kalk:1970,Waldrop:2015}, which provides structural support to the myocardium. Muscle fibers spiral around the heart tube itself, and action potentials propagate to induce myocardial contraction. These action potentials have been previously measured \cite{Kriebel:1967}. Myocardial contractions may begin at either end of the heart tube, allowing the propagation of the action potential to occur in either direction \cite{Anderson:1968}. However, we do not concern ourselves with flow reversals in this model. Although the tunicate heart tube has different material properties and physiological properties than the vertebrate embryonic heart, it still is an interesting model for vertebrate heart morphogenesis \cite{Morad:1980}. However, the conduction properties, e.g., velocities, of action potentials are much more uniform in tunicates than mammalian hearts \cite{Jucker:1968}.

\begin{figure}[H]
    \centering
    \includegraphics[width=0.5\textwidth]{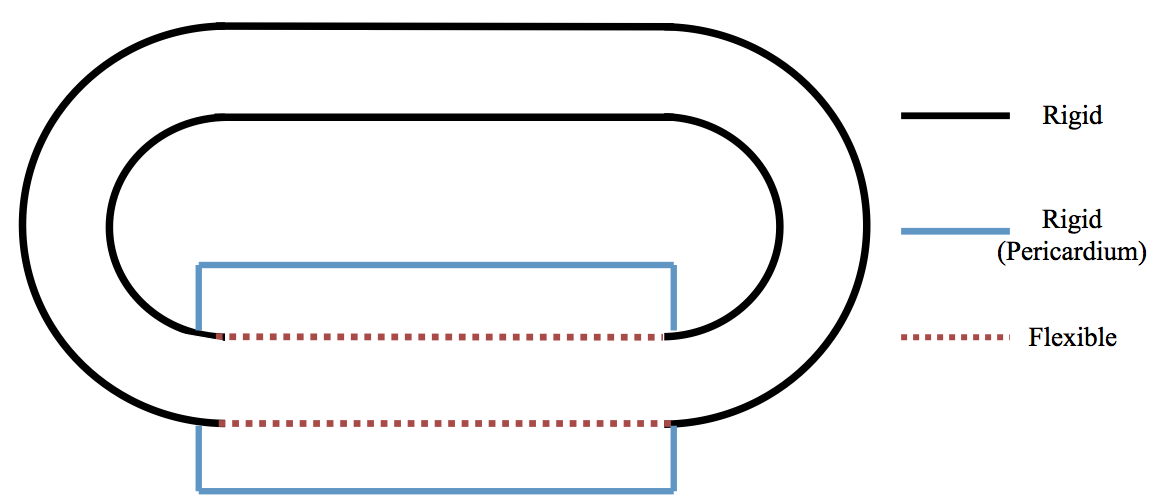}
    \caption{Computational geometry for the electro-mechanical pumping model. The racetrack is held stiff (black), except for the bottom straight-tube portion, which is flexible (red). There is a stiff pericardium model surrounding the flexible region (blue).}
    \label{fig:Model_Geometry_Electro}
\end{figure}

The computational model we investigate is shown in Figure \ref{fig:Model_Geometry_Electro}. Linear springs and beams connect adjacent Lagrangian points in the flexible region of the racetrack geometry. All other Lagrangian points of the boundary are modeled using target points, to hold the stiff portions of the racetrack and pericardium region nearly rigid. The parameters used in the model are found in Table \ref{HT_Electro_Params_1} below.
\begin{table}[H]
\centering
\begin{tabular}{|c|c|}
\hline
{\bf{Parameter}}       & {\bf{Value}} \\
\hline
Length/Width of comp. domain (m) & $5.0\times10^{-4}$ \\
Diameter of tube [$d$] (m) & $3.5\times10^{-5}$ \\
Outer Radius [$R_o$] (m)  & $1\times10^{-5}$ \\
Inner Radius [$R_i$] (m) & $d-R_o$ \\
Length of Straight Tube (m) & $5.0\times10^{-4}$ \\
Eulerian Resolution [dx] (m) & $8.33\times10^{-7}$ \\
Lagrangian Resolution [ds] (m) & $4.17\times10^{-7}$ \\
Density of fluid ($\rho$)$\left[\frac{kg}{m^3}\right]$ & 1025 \\
Viscosity of fluid ($\mu$)$\left[\frac{kg}{ms}\right]$ & \textit{varied} \\
Stretching stiffness of the boundary ($k_{spr}$)$\left[\frac{kg}{s^2}\right]$ & $3.24\times 10^5$ \\
Stretching stiffness of target points ($k_{target}$)[$Nm^2$]  & $3.24\times 10^5$ \\
Bending coefficient of boundary ($k_{beam}$)$\left[\frac{kg}{s^2}\right]$ & $3.24\times 10^5$ \\
\hline
\end{tabular}
\caption{Table of the parameters associated with the fluid and the immersed boundary fiber models.}
\label{HT_Electro_Params_1}
\end{table}
Instead of prescribing contraction, we develop a model for the underlying electrophysiology of the heart, i.e., traveling action potentials arising from a single pacemaker region, to couple to myocardial contraction and hence intracardiac fluid flow. The model of action potential propagation is given by the FitzHugh-Nagumo equations \cite{FitzHugh:1961,Baird:2015} below,
\begin{align}
\label{FHN:eq_1} \frac{\partial v}{\partial t} &= \mathbb{D} \nabla^2 v + v(v-v_a)(v-1) - w - \mathbb{I}(t) \\
\frac{\partial w}{\partial t} &= \epsilon(v-\gamma w),
\label{FHN:eq_2}
\end{align}
where $v(x,t)$ is the membrane potential, $w(x,t)$ is the blocking mechanism, $\mathbb{D}$ is the diffusion rate of the membrane potential, $v_a$ is the threshold potential, $\gamma$ is the resetting rate, $\epsilon$ is the blocking strength parameter, and $\mathbb{I}(t)$ is an applied current, e.g., an initial stimulus potentially from pacemaker signal activation. Note that $v$ is the action potential and that $w$ can be thought to model a sodium blocking channel. We note that the FitzHugh-Nagumo equations (\ref{FHN:eq_1})-(\ref{FHN:eq_2}) are a reduced order model of the Hodgkin-Huxley equations, which were the first quantitative model to describe the propagation of an electrical signal across excitable cells \cite{Hodgkins:1952}. The parameters used in the electrophysiology model are found in Table \ref{HT_Electro_Params_2}.

\begin{table}
\centering
\begin{tabular}{|c|c|}
\hline
{\bf{Parameter}}       & {\bf{Value}} \\
\hline
Threshold potential [$v_a$] & $0.1$ \\
Strength of blocking [$\epsilon$] & $0.1$ \\
Diffusive coefficient [$\mathbb{D}$] & $0.1-100$ \\
Resetting rate [$\gamma$]   & $0.5$ \\
Current injection [$\mathbb{I}$] & $0.5$ \\
Frequency [$f$] (Hz) & $1.0$\\
\hline
\end{tabular}
\caption{Table of the parameters associated with the FitzHugh-Nagumo electrophysiology model.}
\label{HT_Electro_Params_2}
\end{table}
Next we need to interpolate the information from the electrophysiology model to the fluid-structure interaction solver, i.e., immersed boundary method. Time is scaled in order to match the dynamics of the generated action potentials to the desired active wave of contraction and is given by:
\begin{equation}
\label{FHN_dt} dt_f = \frac{dt\mathbb{F}}{\mathbb{T}},
\end{equation}
where $dt$ is the time-step associated with the fluid solver, $\mathbb{F}$ is a non-dimensional scaling parameter, and $\mathbb{T}$ is the desired pumping period. The spatial location, $x$, in (\ref{FHN:eq_1})-(\ref{FHN:eq_2}) is also scaled to match the dynamics of the active wave of contraction on the tube. When the propagating action potential reaches one of the muscles along the tube, the associated spring stiffness of said muscle model is given by
\begin{equation}
\label{FHN_muscle_spring} k_{e}(x,t) = k_m\left( v^4(x,t)\right).
\end{equation}
The simplified muscle model is given by a dynamic spring stiffness coefficient, given by $k_e(x,t)$, which is a non-linear function of the traveling action potential, $v(x,t)$. This idea was adapted from Baird \textit{et al.} \cite{BairdThesis:2014,Baird:2015}. The force generated by the springs that connect the bottom and top of the elastic tube can then be computed. These forces represents muscular contraction. The value of $k_m$ is tuned to produce the amount of contraction observed in Ciona hearts, as in \cite{BairdThesis:2014,Baird:2015}.

\begin{figure}[H]
    \centering
    \includegraphics[width=0.5\textwidth]{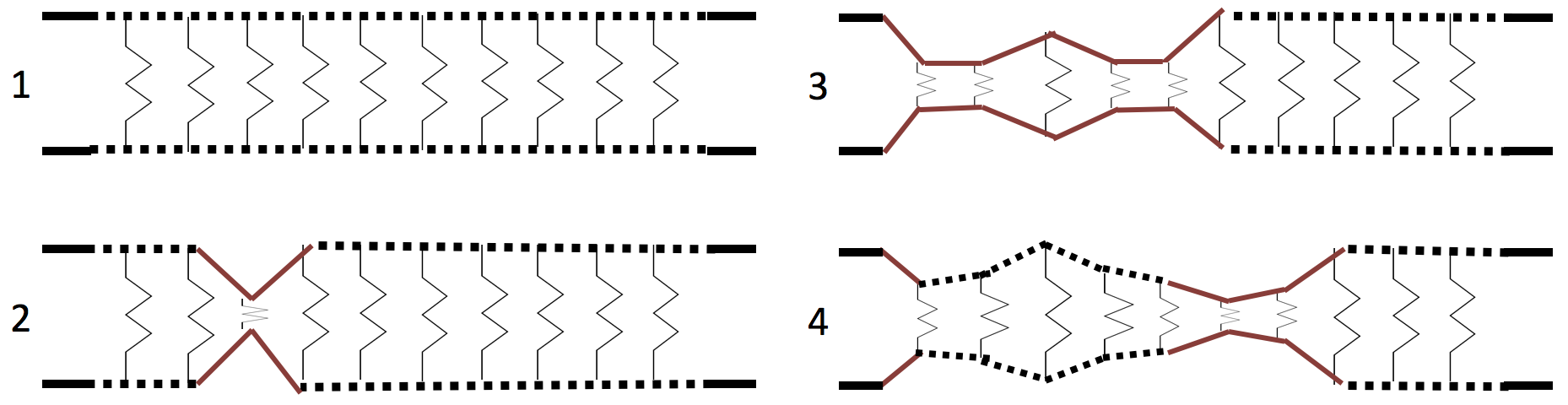}
    \caption{Schematic of electrodynamical pumping. (1) The tube at rest; the springs connecting the top and bottom of the tube are the muscles. (2) The pacemaker initates an action potential, in which the tube will contract based on the magnitude of the signal (3)-(4) The action potential propagates along the tube, induing contraction.}
\label{fig:electro_schematic}
\end{figure}

The idea for electro-dynamic pumping can be seen in Figure \ref{fig:electro_schematic}, which is a schematic for electro-dynamical pumping behavior. First the tube is at rest until a pacemaker initiates a potential signal, which contracts the tube in one singular region. Next the action potential propagates along the tube inducing contraction. Once the action potential passes outside a region on the tube, that location no longer has active contraction, but can return to its resting position. 

\begin{figure}[H]
    \centering
    \includegraphics[width=0.5\textwidth]{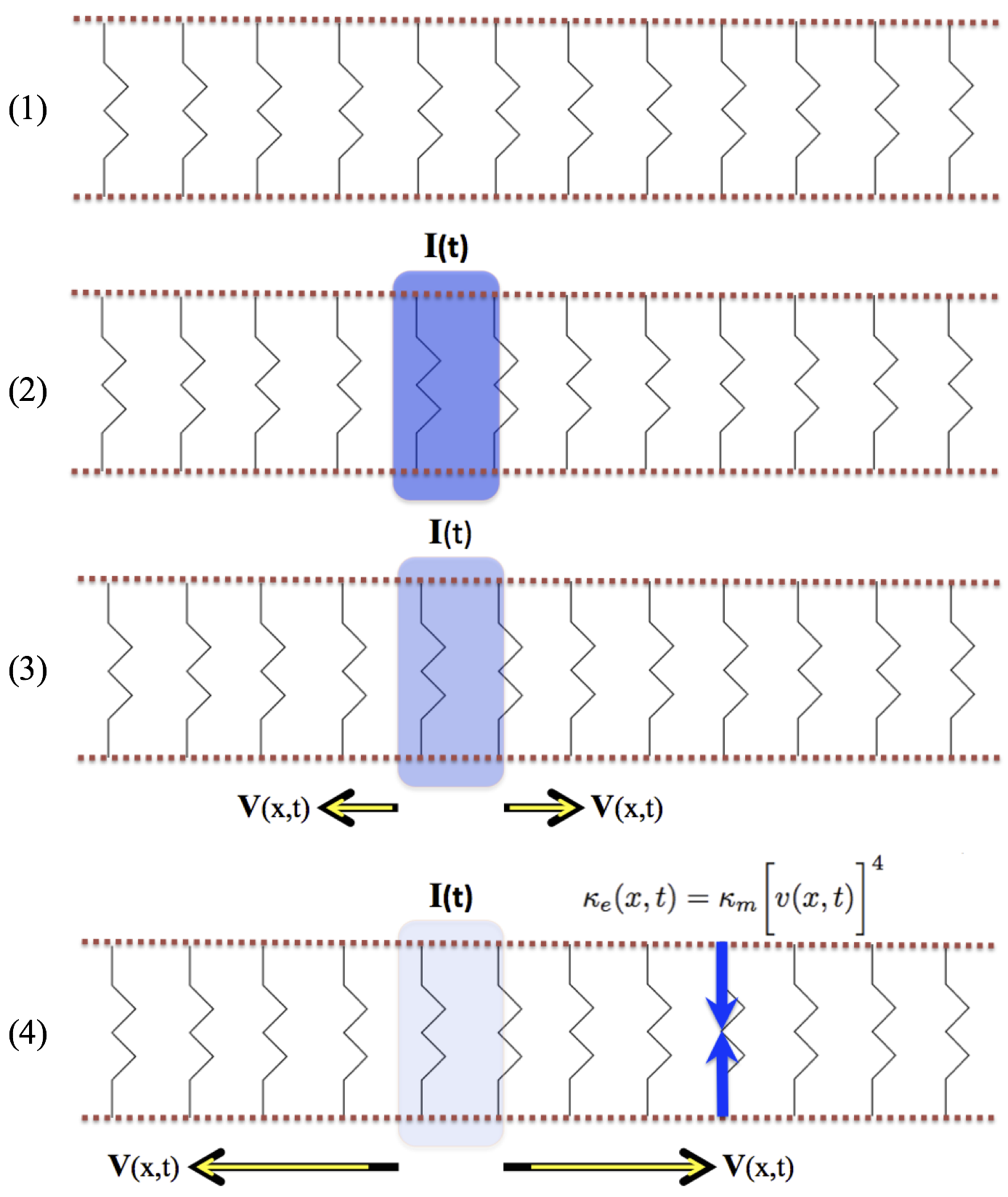}
    \caption{Schematic of electrodynamical pumping. (1) The tube at rest; the springs connecting the top and bottom of the tube are the muscles. (2) The pacemaker initates an action potential, in which the tube will contract based on the magnitude of the signal (3)-(4) The action potential propagates along the tube, induing contraction. }
\label{fig:electro_schematic2}
\end{figure}

Furthermore the main electrophysiology idea behind the model is illustrated in Figure \ref{fig:electro_schematic2}. In diagram 1 the flexible tube is at rest. Next 2 depicts a pacemaker initiating an input signal (current). Then that voltage (action potential) travels down the tube, while the input signal dissipates. Once the action potential reaches a muscle fiber, the tube contracts based on a non-linear relationship between spring stiffness and the magnitude of the action potential (voltage).

%
%
\section{Results}

In this study, we conducted numerical experiments of the electro-dynamic pumping model, which encompassed fully coupled electrophysiology to pumping behavior for a heart tube, modeled as a closed racetrack geometry. We investigated various diffusivities, $\mathbb{D}$, which give rise to different pumping regimes, e.g., either a `dynamic suction pumping-esque' or `peristaltic-like' pumping regime. Furthermore, we explored these regimes for over $3$ orders of magnitude in $Wo$.

%
%

\subsection{Results of the FitzHugh-Nagumo Model}

Here we present the varying action potential dynamics given via the FitzHugh-Nagumo equation, which models the electrophysiology. We explored this model for a variety of diffusive coefficients, $\mathbb{D}=\{0.1,1.0,10.0,100.0\}$.

\begin{figure}[H]
    \centering
    \includegraphics[width=0.5\textwidth]{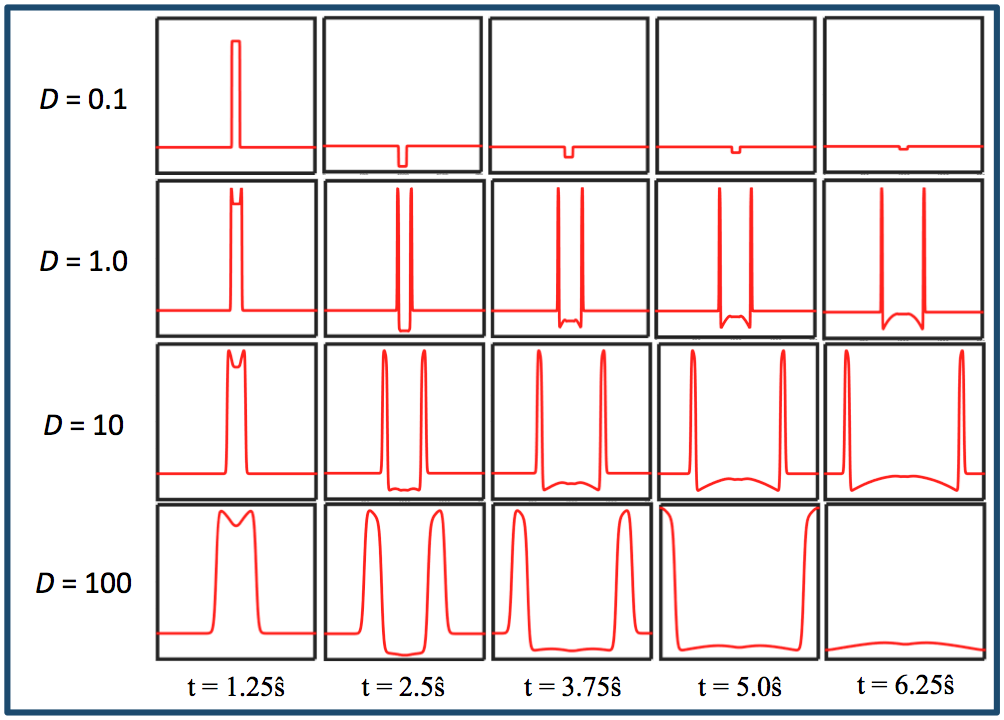}
    \caption{Different traveling wave propagation properties arising out of the FitzHugh-Nagumo equations for varying diffusivities, $\mathbb{D}=\{0.1,1.0,10.0,100.0\}$.}
\label{fig:electro_FHN_wave}
\end{figure}

Figure \ref{fig:electro_FHN_wave} illustrates the kinds of traveling action potentials that arise out of the electrophysiology model. It is clear that the $\mathbb{D}=0.1$ case resembles a signal that could be reminiscent of that of dynamic suction pumping, where as $\mathbb{D}$ gives rise to a propagating action potential that could model a more peristaltic-like contraction. It is clear that as diffusivity increases, the waves propagate outwards, and with greater wave-speed. Furthermore, the wave-form itself gets wider. 

%
%

\subsection{Results of the electro-dynamical heart tube model}

In this section we present the results describing how bulk flow rates are affected by varying the diffusivity, to capture different pumping behaviors for a variety of $Wo$. 

\begin{figure}[H]
    \centering
    \includegraphics[width=0.5\textwidth]{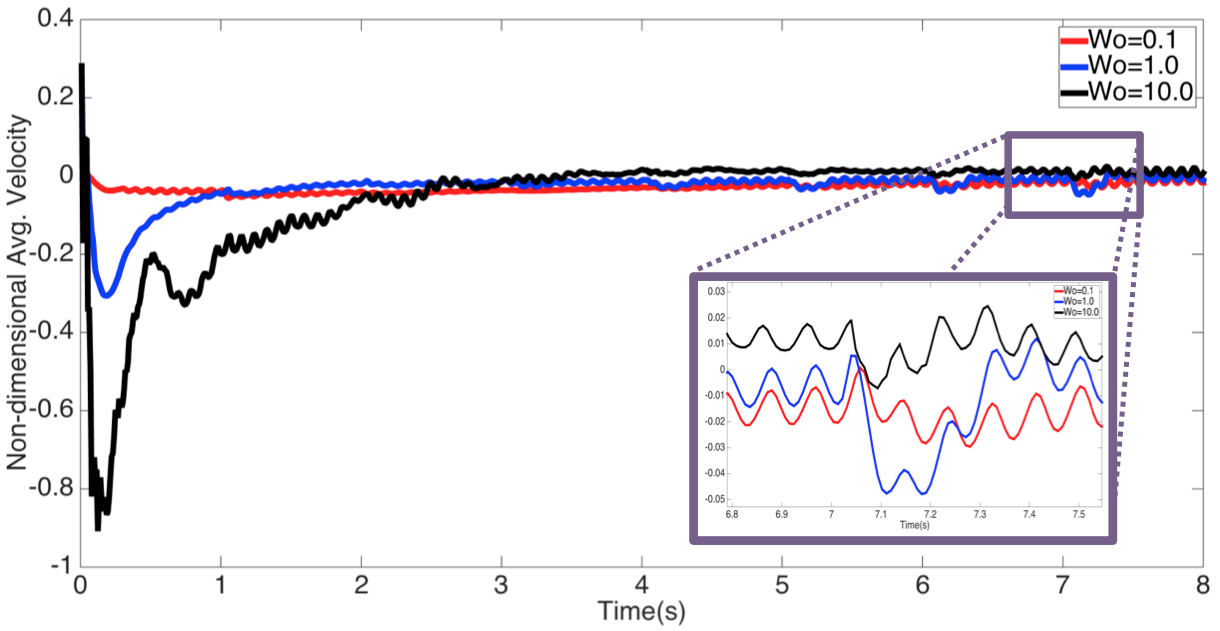}
    \caption{The non-dimensional spatially-averaged velocity computed across a cross-section of the top of the race-track geometry vs non-dimensional time for $\mathbb{D}=0.1$, e.g., the `dynamic suction pumping' regime, for $Wo=\{0.1,1.0,10.0\}$. The zoomed in portion illustrates the resulting wave-form and the high frequency oscillations that result from this pumping regime.}
\label{fig:electro_d0pt1_WoSweep}
\end{figure}

\begin{figure}[H]
    \centering
    \includegraphics[width=0.5\textwidth]{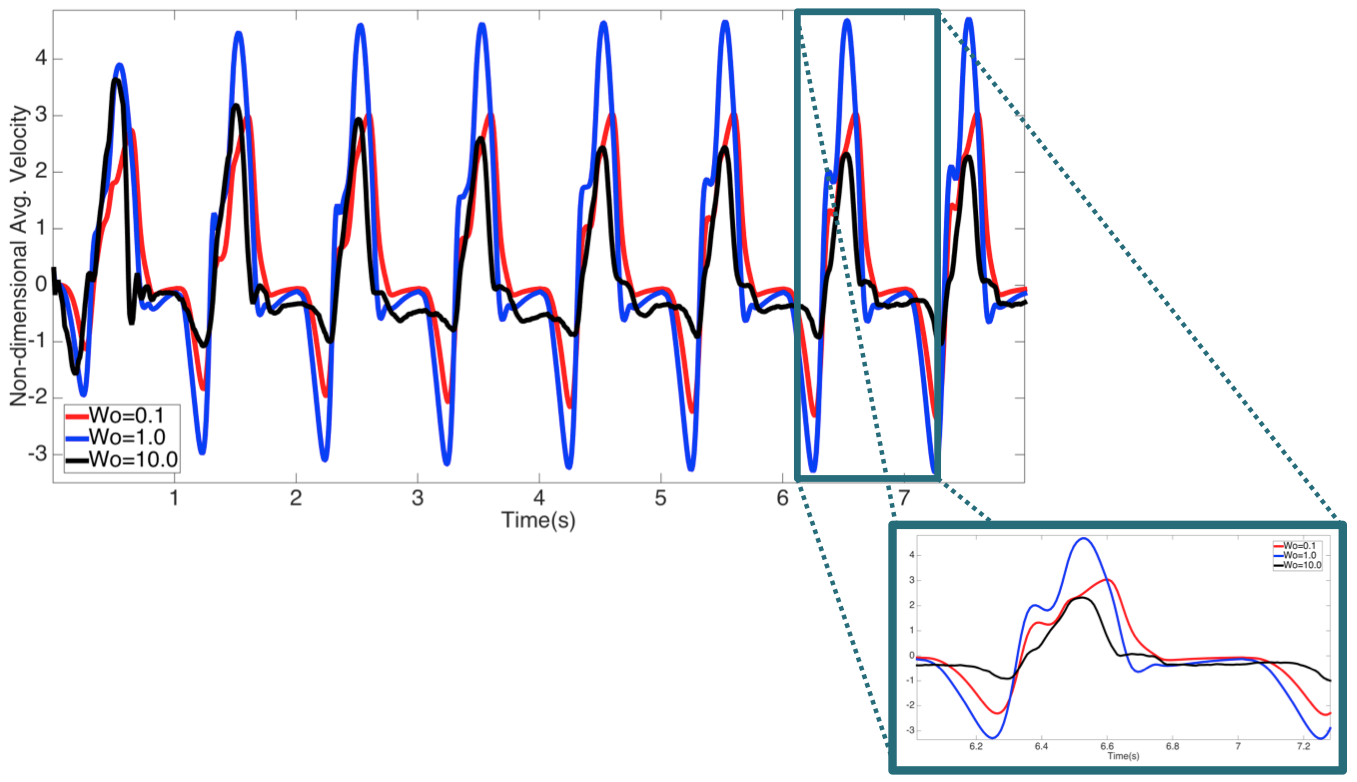}
    \caption{The non-dimensional spatially-averaged velocity computed across a cross-section of the top of the race-track geometry vs non-dimensional time for $\mathbb{D}=100.0$, e.g., the `peristaltic' regime, for $Wo=\{0.1,1.0,10.0\}$. The zoomed in portion illustrates the resulting wave-form.}
\label{fig:electro_d100_WoSweep}
\end{figure}

Figures \ref{fig:electro_d0pt1_WoSweep} and \ref{fig:electro_d100_WoSweep} illustrate the non-dimensional spatially-averaged velocity computed across a cross-section of the top of the race-track geometry vs non-dimensional time for $\mathbb{D}=0.1$ (Figure \ref{fig:electro_d0pt1_WoSweep}) and $\mathbb{D}=100.0$ (Figure \ref{fig:electro_d100_WoSweep}). It is clear that when $\mathbb{D}=0.1$ there is not significant bulk flow produced regardless of $Wo$, unlike the case when $\mathbb{D}=100.0$, where significant bulk flow is produced over all $Wo=\{0.1,10,10\}$. It is also clear that the wave-form produced for $\mathbb{D}=0.1$ undergoes many more high frequency oscillations as compared to the case for $\mathbb{D}=100$.

\begin{figure}
    \begin{center}
    \begin{subfigure}[b]{0.49\textwidth}
        \centering
        \includegraphics[width=\textwidth]{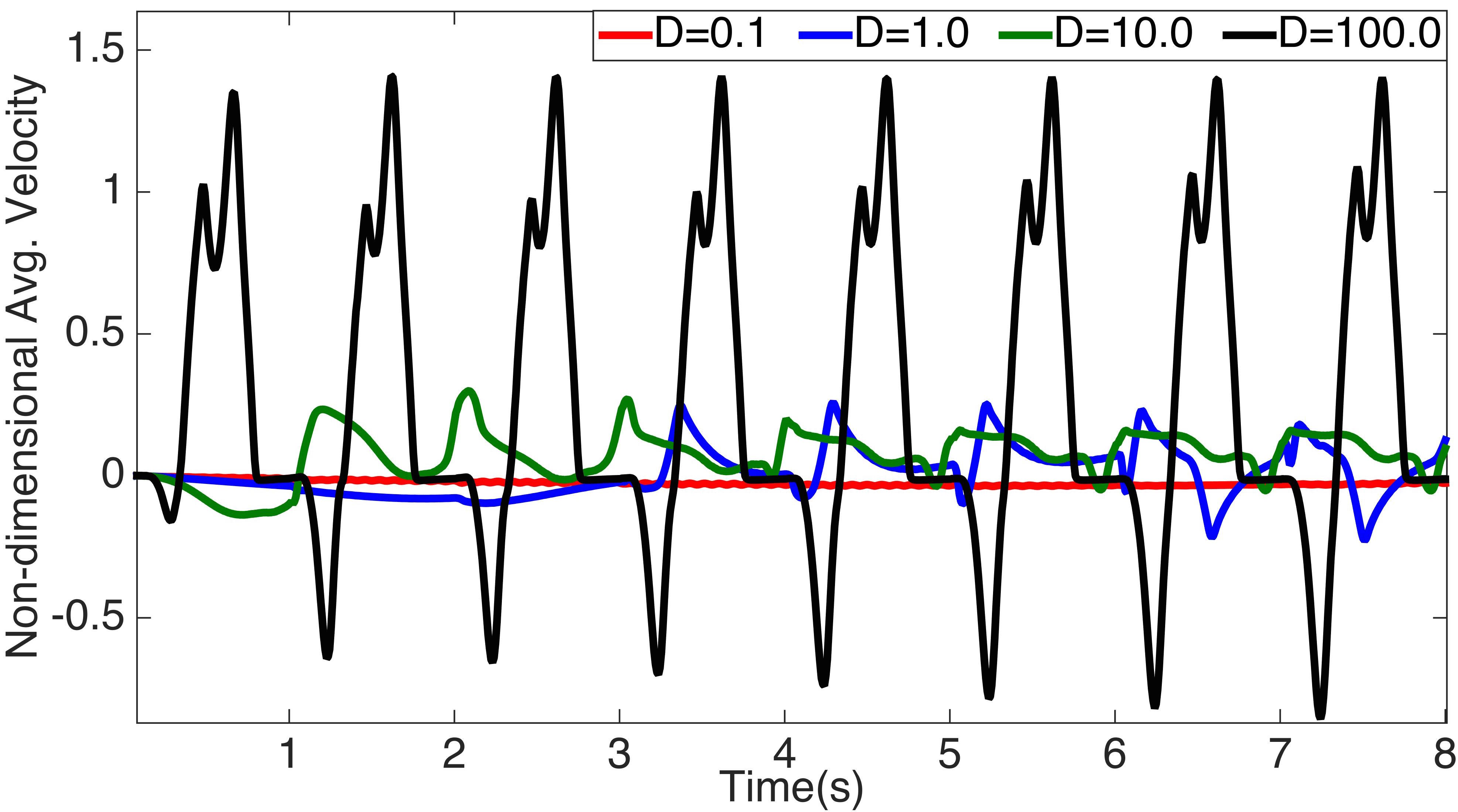} \hfill
        \caption{}
        \label{fig:HT_Electro_Wo0pt1_diff}
    \end{subfigure} \\
    \end{center}
    \begin{center}
    \begin{subfigure}[b]{0.49\textwidth}
        \centering
        \includegraphics[width=\textwidth]{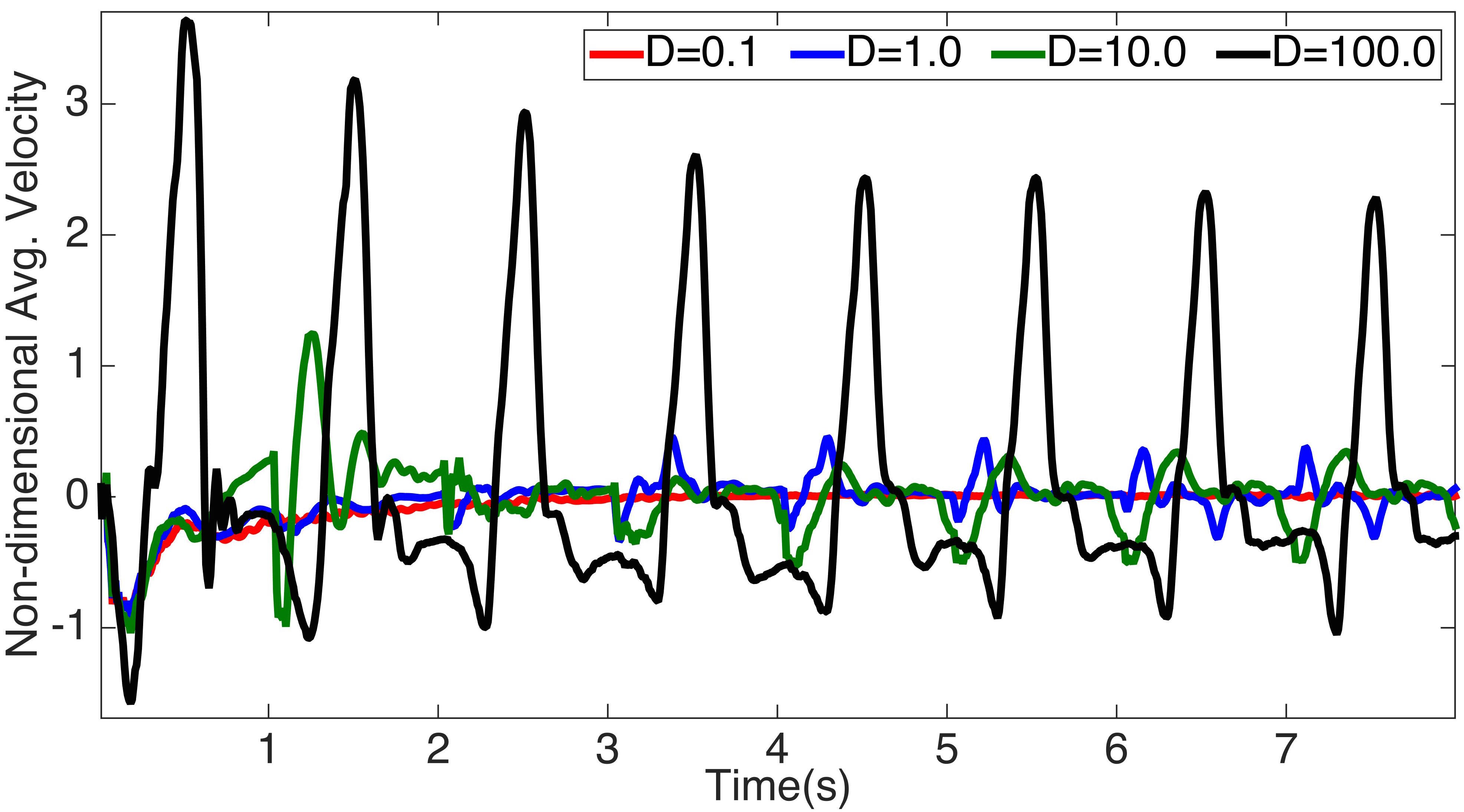} \hfill
        \caption{}
        \label{fig:HT_Electro_Wo10_diff}
    \end{subfigure}\\
    \end{center}
\caption{A comparison of non-dimensional spatially-averaged velocity computed across a cross-section at the top of the racetrack vs non-dimensional time in the simulation for varying diffusive coefficients, $\mathbb{D}=\{0.1,1.0,10.0,100.0\}$. The two plots compare different $Wo$, e.g., (a) $Wo=0.1$ and (b) $Wo=10$. }
\label{HT_Electro_Wo_diff}
\end{figure}

Comparing corresponding $Wo$ pumping mechanisms for a variety of $\mathbb{D}=\{0.1,1.0,10.0,100.0\}$ are shown in Figure \ref{HT_Electro_Wo_diff}, where Figure \ref{fig:HT_Electro_Wo0pt1_diff} compares pumping regimes for $Wo=0.1$ and Figure \ref{fig:HT_Electro_Wo10_diff} for $Wo=10$. It is clear that in both cases that the most bulk flow is produced when $\mathbb{D}=100$, and some flow is produced in the cases of $\mathbb{D}=\{1,10\}$. There is still backflow in the $\mathbb{D}=100$ case and less overall backflow in the $\mathbb{D}=10$ case.

Furthermore, the wave-form in the $\mathbb{D}=100$ case is different between the $Wo=0.1$ and $Wo=10$ cases. There is a single peak for the case when $Wo=10$ and a dual peaks for $Wo=0.1$ for the forward flow; however, in the backflow, the situation is reversed, where a dual-peak is observed for $Wo=10$ and a single peak for $Wo=0.1$. 

\begin{figure}
    \centering
    \includegraphics[width=0.5\textwidth]{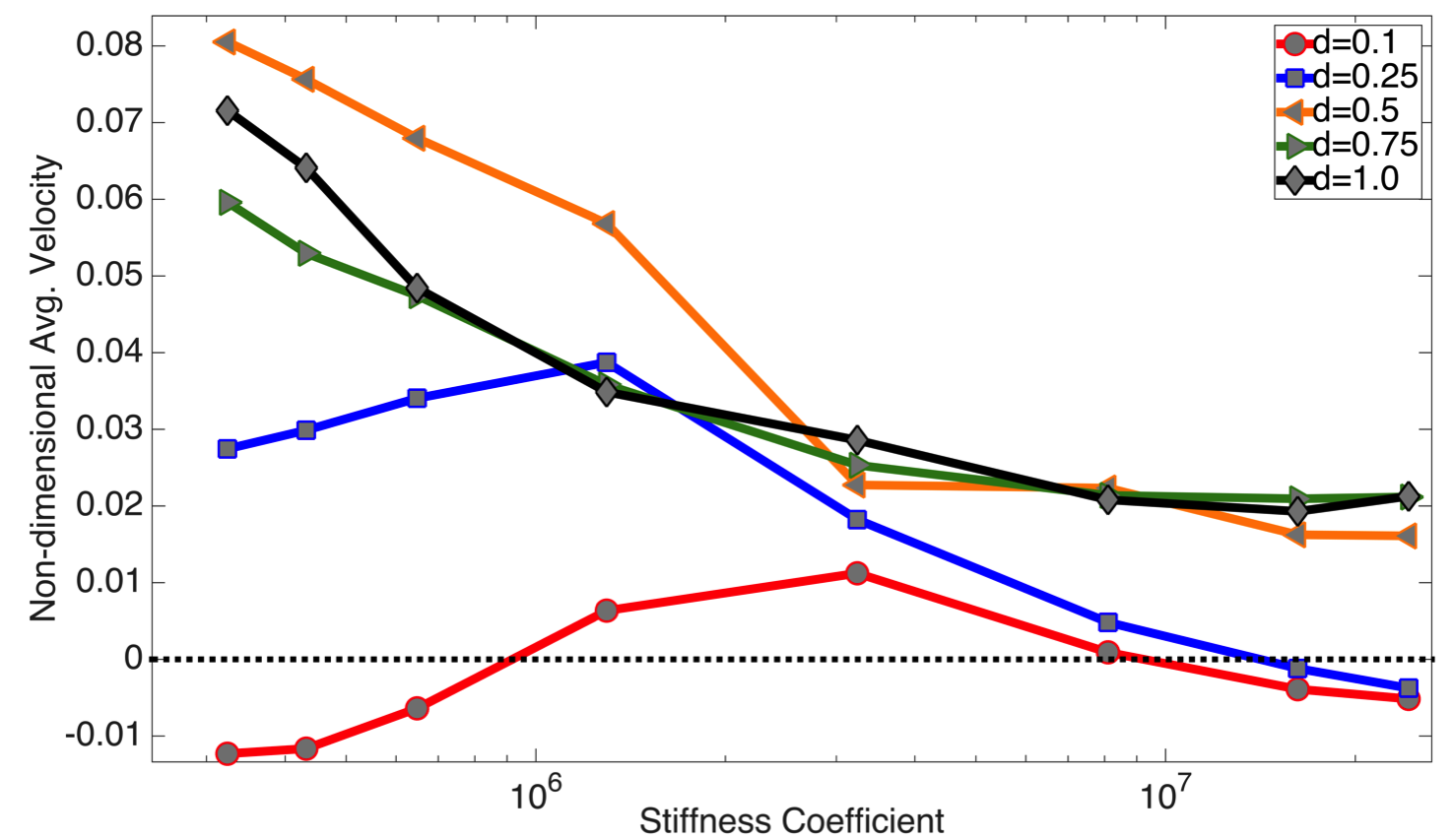}
    \caption{A plot of non-dimensional spatially-averaged velocity computed across a cross-section at the top of the racetrack vs the non-dimensional stretching and bending stiffness coefficients for pumping in the `dynamic suction pumping' regime, for a variety of diffusivities, $\mathbb{D}=\{0.1,0.25,0.5,0.75,1.0\}$.}
\label{fig:electro_AvgV_Stiffness}
\end{figure}

In attempt to maximize bulk flow for the dynamic suction pumping-esque regime, the stretching-stiffness and bending stiffness coefficients of the tube were varied. The results are shown in Figure \ref{fig:electro_AvgV_Stiffness}. It is clear that as the stiffness is varied there is a non-linear relationships between flow speed (spatially- and temporally-averaged non-dimensional velocity across a cross-section of the racetrack) and stiffness. However, not a considerable amount of more bulk flow was produced from increasing these stiffness coefficients. 

\begin{figure}
    \centering
    \includegraphics[width=0.5\textwidth]{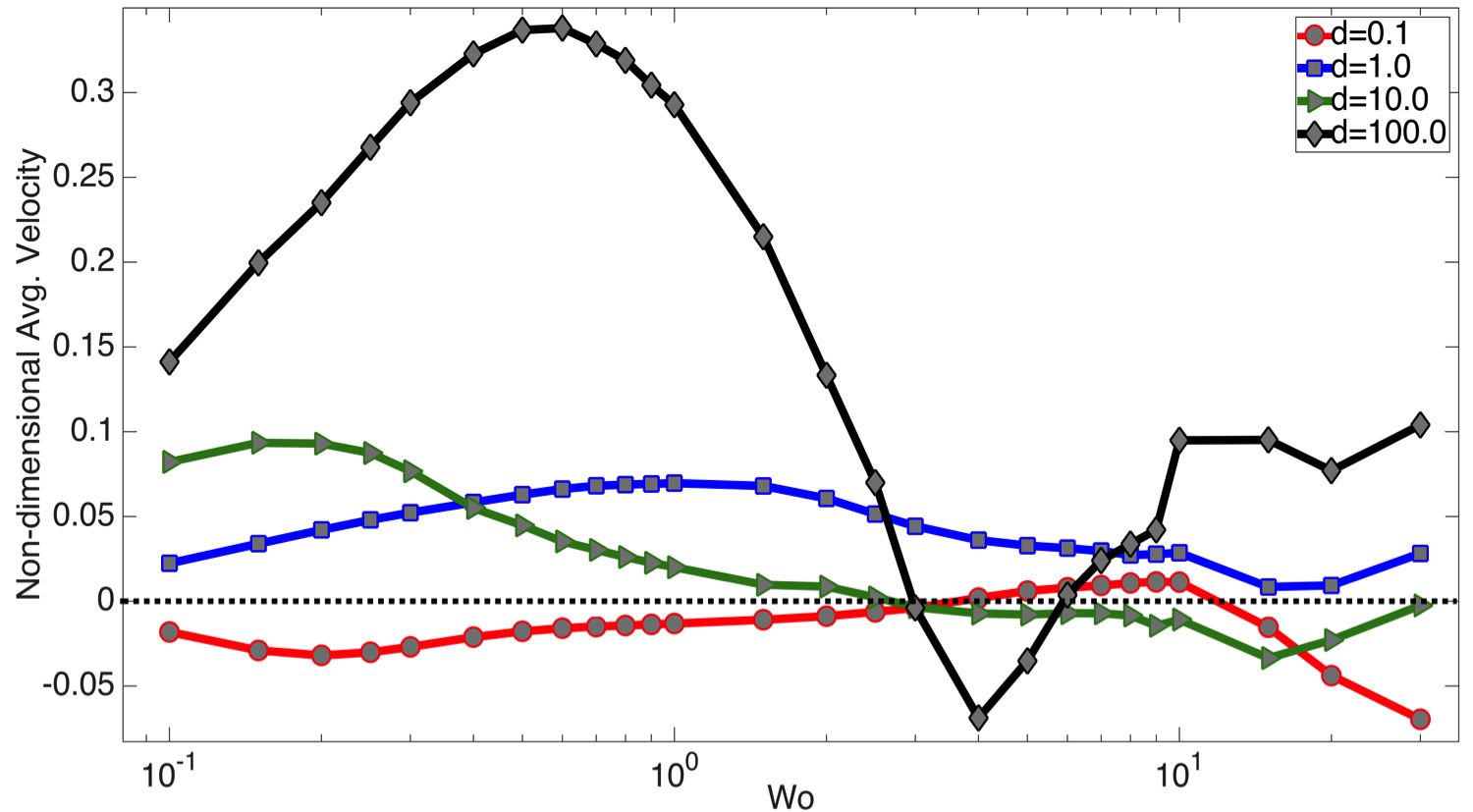}
    \caption{A comparison of the spatially- and temporally-averaged non-dimensional velocities computed across a cross-section of the racetrack vs. $Wo$ for varying diffusivities, $\mathbb{D}=\{0.1,1.0,10.0,100.0\}$.}
\label{fig:electro_AvgV_WoSweep}
\end{figure}

Lastly we compared the spatially- and temporally-averaged non-dimensional velocities across a cross-section of the racetrack against $Wo$ for a variety of $\mathbb{D}$. The results are shown in Figure \ref{fig:electro_AvgV_WoSweep}. It is clear there is a non-linear relationship in average velocity and scale arising from this model of pumping in every pumping regime, given by $\mathbb{D}$. Furthermore, the highest bulk flow rates were seen in the case of $\mathbb{D}=100$ for $Wo\sim 0.8$, which correspond to the $Wo$ around that of tunicate tubular hearts \cite{Baird:2014,BairdThesis:2014}.  

%
%

\section{Discussion and Conclusion}

This $2D$ model coupled the propagation of action potentials, given via the FitzHugh-Nagumo equations, to the force generation and myocardial contraction, given through a non-linear spring-like muscle model, to induce pumping behavior in a flexible tube, where the fully coupled fluid-structure interaction model was solved using the immersed boundary method. This model was first described in \cite{Baird:2015}. We explored the effect of perturbing a diffusive coefficient in the electrophysiology model to capture different pumping regimes. 

It was clear that by varying this diffusive term, $\mathbb{D}$, the model was able to recreate a spectrum of pumping mechanisms, ranging from one that in which the action potential remained localized and did not diffusive, i.e., a dynamic suction pumping-esque behavior, and one where the action potential diffused along the heart tube in as a more traveling wave, e.g., peristaltic-like active wave of contraction. Our model showed that when $\mathbb{D}$ was in the more peristaltic-like regime, i.e., $\mathbb{D}\sim 100$, more bulk flow was produced in the racetrack geometry, as compared to more negligible amounts from the dynamic suction pumping-esque regime, $\mathbb{D}\sim 0.1.$ This result was consistent for the range of $Wo$ considered.  

Moreover, in all cases considered, there was a non-linear relationship between average flow rate, scale ($Wo$), and diffusivity (pumping behavior). More bulk flow was produced on average (both spatially and temporally), with a maximum around $Wo\sim 0.8$ than for higher $Wo$, up to $Wo=30$, in the peristalic-like regime. Similar behavior, in that peristalsis produces more bulk flow than DSP, has been observed before when using prescribed pumping behavior, as in \cite{Baird:2014,Battista:2016a}. 

However, perturbing the material properties of the tube could potentially affect bulk flow rates across all pumping regimes, given by $\mathbb{D}.$ Our focus was limited to perturbing the stretching and bending stiffnesses of the tube specifically within the dynamic suction pumping-esque regime, $\mathbb{D}\sim[0.1,1]$. Furthermore, our study only considered increasing the stiffnesses and not decreasing them. For the regime and material properties considered, we found a non-linear relationship between flow rates and stiffness. 

As blood flow and the resulting hemodynamic forces are essential for proper heart development \cite{Hove:2003}, it is important that the pumping model capture as much biology as possible. Each pumping regime considered here, given by the diffusivity of action potential propagation, will give rise to a different force distribution along the endothelial lining of the heart and hence impact the epigenetic signals that are transmitted through mechanotransduction \cite{French:1992,Weckstrom:2007}. Furthermore the flow profiles resulting from each pumping mechanism would be different. These differences in the flow patterns itself could impact the way morphogens advect and diffuse during embryogenesis \cite{Gurdon:2001,Christian:2012}, opening the realm to a lot more interesting biological questions to explore.

%
%

\section*{Acknowledgment}

The authors would like to thank Steven Vogel for past conversations on scaling in various hearts as well as Lindsay Waldrop, Austin Baird, Julia Samson, and William Kier for discussions on embryonic hearts. They also wish to thank the organizers of the 2017 BIOMATH Conference in Kruger Park, South Africa. This project was funded by NSF DMS CAREER \#1151478, NSF CBET \#1511427, NSF DMS \#1151478, NSF POLS \#1505061, NSF PHY \#1504777, and NSF DMS \#1127914  awarded to L.A.M. Funding for N.A.B. was provided from an National Institutes of Health T32 grant [HL069768-14; PI, Christopher Mack] and UNC Dissertation Completion Fellowship.

%
%

\bibliographystyle{elsarticle-num}
\bibliography{heart}

%
%

\end{document}